\documentclass[11p,twocolumn]{revtex4-2}

\usepackage{mathrsfs}
\usepackage{amsfonts}
\usepackage{amssymb}
\usepackage{amsmath}
\usepackage{graphicx}
\usepackage{color}
\usepackage[ruled,vlined]{algorithm2e}
\usepackage{algorithmic}
\usepackage{stackrel}
\usepackage{listings}
\usepackage{enumitem}

\newcommand{\Tr}{\mathrm{Tr}}

\newcommand{\be}{\begin{equation}}
\newcommand{\ee}{\end{equation}}
\newcommand{\bqa}{\begin{eqnarray}}
\newcommand{\eqa}{\end{eqnarray}}

\graphicspath{{./figures/}}

\begin{document}

\title{A Hardware Accelerator for the Goemans-Williamson Algorithm}

\author{D. A. Herrera-Mart\'i}
\author{E. Guthmuller}
\author{J. Fereyre}
\affiliation{Universit\'e Grenoble Alpes, CEA List, 38000 Grenoble, France}

\date{\today}

\begin{abstract} 

The combinatorial problem Max-Cut has become a benchmark in the evaluation of local search heuristics for both quantum and classical optimisers. In contrast to local search, which only provides average-case performance guarantees, the convex semidefinite relaxation of Max-Cut by Goemans and Williamson, provides worst-case guarantees and is therefore suited to both the construction of benchmarks and in applications to performance-critic scenarios.

We show how extended floating point precision can be incorporated in algebraic subroutines in convex optimisation, namely in indirect matrix inversion methods like Conjugate Gradient, which are used in Interior Point Methods in the case of very large problem sizes. Also, an estimate is provided of the expected acceleration of the time to solution for a hardware architecture that runs natively on extended precision. Specifically, when using indirect matrix inversion methods like Conjugate Gradient, which have lower complexity than direct methods and are therefore used in very large problems, we see that increasing the internal working precision reduces the time to solution by a factor that increases with the system size.
\end{abstract}
\maketitle

\section{Introduction}

The space of solutions in combinatorial optimisation problems grows exponentially with the size of the instances, making exhaustive search infeasible. Some systematic methods explore the solution space and provide exact solutions, but their complexity is exponential in the problem size \cite{korte08}.  Local search heuristics provide approximate solutions when exact methods are computationally prohibitive. These algorithms start from an initial solution and iteratively move to neighbouring candidates to improve quality \cite{gendreau10}. 

Physics-inspired local search heuristics include simulated annealing, quantum annealing, simulated quantum annealing, coherent Ising machines and simulated bifurcation machines \cite{morita08, crosson16, hauke20, inagaki16, yamamoto17, yamamotok17} among others. Although they do not guarantee optimality in finite time, these methods are routinely used to find high quality solutions for large unconstrained quadratic binary problems. From FPGAs and GPUs to dedicated quantum and quantum-inspired processors, hardware acceleration has recently become a key strategy to improve performance of these methods \cite{minamisawa19,cook19,tatsumura19, tatsumura21, sao19, mondal20, honjo21, mohseni22, nikhar24}

The Max-Cut problem is a classical NP-hard combinatorial problem that is routinely used as a benchmark for classical and quantum optimisers. The  Goemans-Williamson's (GW) algorithm provides an approximation ratio of ca. $0.879$ in the worst case \cite{GW94,GW95} for this problem. The GW algorithm achieves this by relaxing the binary constraints of the Max-Cut and reformulating it as an semidefinite program (SDP). This relaxation is typically solved via Interior Point Methods (IPM), which are second-order optimisation methods relying on matrix inversion, either via direct numerical algebra, or via indirect Krylov methods, such as conjugate gradient (CG), in the limit of very large problems. Krylov methods, and CG in particular, are severely limited by machine precision, and one of the goals of this paper is to show how increasing floating point precision results in less iterations for indirect matrix inversion. 

It has been shown \cite{bravyi20} that the (local) quantum approximate optimisation algorithm (QAOA) often performs worse than the GW algorithm for Max-Cut problems. This raises important questions about the practical quantum advantage in real-world optimisation problems and also highlights the importance of classical algorithms in the construction of benchmarks for quantum or quantum-inspired optimisation algorithms. A second goal of this work is therefore to provide a scalable method to obtain good-enough solutions to SDP relaxations of combinatorial problems (with worst-case guarantees) that could be used as benchmarks for quantum heuristics.

This paper is structured as follows. In the next section, we provide the mathematical framework for our numerical experiments. We explain why, in some cases, increasing the internal working precision can accelerate the convergence of important numerical algebra subroutines, such as CG. We then touch on some considerations about how peak performance is achieved for different kinds of computational tasks, and how this is relevant for SDP relaxations and for combinatorial optimisation in general. We conclude by presenting the results and analysis of our numerical simulations.

\section{Semidefinite Relaxation of Max-Cut}

The Max-Cut problem consists in trying to find a partition in a graph that maximises the number of weights across the two disjoint subsets of vertices. It can be expressed mathematically as an optimisation problem:

\[
\max_{x \in \{ -1, +1 \}^n} \quad  \sum_{i,j=1}^n c_{ij} (1 - x_i x_j).
\]

If one defines a weight matrix $C = (c_{ij}) $ and a vector of assignments $x = (x_1, \ldots, x_n)^T $, the objective can then be rewritten as:

\[
   \sum_{i,j=1}^n c_{ij} - x^T C x .
\]

This is equivalent to minimising the following cost function:

\[
\min_{x \in \{ -1, +1 \}^n} x^T C x.
\]

Notice that $x^T C x = \mathrm{Tr}(C x x^T)$, and that $X = x x^T $ is a rank-1 symmetric positive semidefinite matrix such that $X_{ii} = x_i^2 = 1$.

The hardness of Max-Cut can be seen from the binary, non-convex, constraints of this matrix form. The main idea behind the GW algorithm is to promote binary variables to vectors, such that the condition $x_i x_j = \pm 1$ corresponds to vectors being parallel or anti-parallel. The next step of the relaxation, which renders the problem convex and therefore tractable, is to allow the vectors to take values in a continuum of values rather than in just a discrete set. 

Dropping the rank constraint yields the \emph{semidefinite relaxation}:

\[
\begin{aligned}
\min_{X} \quad & \mathrm{Tr}(C X) \\
\text{subject to} \quad & X_{ii} = 1, \quad i = 1, \ldots n \\
& X \succeq 0
\end{aligned}
\]

Which can be solved using standard techniques in convex optimisation for SDPs, such as Interior Point Methods (IPM). After solving the SDP, a random rounding procedure converts the continuous solution back into a binary cut. Geometric arguments allow one to obtain an approximation ratio of $0.879$ . The Unique Games Conjecture, proposed by Khot in 2002 and which has not been itself proved, suggests that the GW relaxation is optimal unless $P=NP$ \cite{trevisan12}. This technique has been successfully generalised to other combinatorial problems, such as Max-2SAT, Max-DICUT and various constraint satisfaction problems \cite{GW94,feige95,GW01}

\subsection{Interior Point Methods for GW Algorithm}

IPMs solve SDPs by traversing the interior of the feasible region along a central path towards an optimal solution \cite{alizadeh95,boyd04}. The initial point is typically an infeasible point belonging to the cone of positive definite matrices. We implemented a primal-dual barrier method, which simultaneously solves the primal and dual problems while monitoring a duality gap which serves as stopping criterion(in our case it was always $\epsilon=0.005$ in absolute terms). As the optimisation proceeds and the barrier is reduced, the tentative solutions get closer to the feasible set. 

In order to move along the central path, each Newton step has to be computed while respecting some conditions encoded in a quadratic program (see Appendix). Solving the Karush-Kuhn-Tucker (KKT) conditions for this program involves inverting a matrix for which, very often, direct approaches such as Cholesky factorisation can be used. For large SDPs derived from Max-Cut relaxations, direct factorisation complexity grows as $O(n^3)$. Conjugate gradient (CG) methods offer a practical alternative by solving Newton's system iteratively as it involves a quadratic complexity and limited memory footprint.

There are two phenomena that have an important impact in the cost of the optimisation. The first one is that, as the tentative solution moves along the central path towards the feasible set, its eigenvalues start to become more dispersed. At the feasible set, the solution is often rank-deficient, i.e. $det(X) = 0$, so getting close to the optimality involves a rapid worsening in the conditioning of the KKT equations, and each matrix inversion becomes more computationally challenging. This can be seen from the fact that that the KKT equations, used to determine the Newton step, depend on $X^{-1}$, so they will be very sensitive to perturbations across given directions as they get close to the boundary (see appendix and pseudocode description of the algorithm, in particular the inversion of M, which depends on X). Another important phenomenon is the densification of the KKT matrix, which happens very early on in the optimisation. This indicates that preconditioning might not be a saving strategy since in the absence of structure and for dense matrices, preconditioning strategies are not guaranteed to help. Because of this fact, we cannot leverage the full power of CG, for which it is not necessary to store the matrix in dense form. Incidentally, this will also thwart sparse Cholesky factorisation.

\section{Performance Considerations}

One important consideration is that the GW algorithm offers a constant approximation ratio for all problem instances, as opposed to local-search heuristics, which typically operate on average-case guarantees. This means that, whereas the approximation ratio of local-search heuristics can surpass the GW bound of $0.878$, these methods can fail catastrophically for some instances of the problem. Worst-case guarantees entail that the approximation ratio will be the GW bound on average, for all instances. Worst-case guaranteed solutions are important in many applications, such as model predictive control, power-plant planning, transportation, energy routing, etc... In the context of quantum benchmarks, worst-case guarantees provide solid baselines for performance of quantum heuristics. 

Another aspect is that local search heuristics typically struggle with hard constraints. The usual approach is to include them as soft constraints, that is, it includes penalty terms that allow the heuristics to aggressively explore the space of solutions, at the expense of sampling unfeasible configurations. Therefore, enforcing hard constraints in algorithms like simulated annealing, quantum annealing or simulated bifurcation machines can be laborious and demands expert control of penalty terms. In SDP approaches, feasibility is explicitly maintained throughout the optimisation process, ensuring that hard constraints are never violated. But it also offers the possibility of hard-coding constraints into the algorithm, which cannot be done in local-search heuristics. Problems with hard constraints will generally benefit from the mathematical rigour of SDPs, while unconstrained problems might be better suited for point-based local search methods.

\SetAlCapHSkip{0pt} 
\SetAlCapFnt{\normalfont} 

\begin{algorithm}[h]
\caption{Primal-Dual Method for SDP}
\begin{algorithmic}[1]
\REQUIRE Matrices $C$, $\{A_i\}_{i=1}^n$, vector $b$, SDP tolerance $\mathrm{tol}_{SDP}$, CG tolerance $\mathrm{tol}_{CG}$, barrier $\theta$, damping $\eta$,  floating point precision $L$, maximum number of iterations $\mathrm{max\_iter}$
\ENSURE Approximate primal-dual solution $(X, y, S)$

\STATE Initialise in central path $X_0 \succ 0$, $S_0 \succ 0$, $y_0$, set $\mu_0 = \frac{\theta}{n}$

\FOR{$k = 0$ to $\mathrm{max\_iter}$}
    \STATE Compute residuals:
    \[
    \begin{cases}
    r_p = b - \big[\langle A_i, X_k \rangle\big]_{i=1}^n \\
    r_d = C - \sum_{i=1}^n y_{i, k} A_i - S_k \\
    e_g = \Tr (C X_k) - b y_k
    \end{cases}
    \]

    \STATE Linearisation the KKT equations for Newton step.
Vectorise and write normal equations :
    \[
    \begin{cases}
    M^{(k)}_{ij} = \Tr (A_i X_k A_j X_k) \\
    \mathrm{rhs}^{(k)}_i = \Tr(A_i  X_k  C  X_k) - \mu_k \Tr(A_i  X_k) 
    \end{cases}
    \]

    \STATE Solve for $y_{k+1}$ :
    \[
    y_{k+1} = CG (M^{(k)}, \mathrm{rhs}^{(k)}, L, \mathrm{tol}_{CG})
    \]

    \STATE Compute Newton Direction $D$:
    \[
    \begin{cases}
    Z = C - \sum_{i=1}^n A_i  y_{i, k+1} \\
    D  = X_{k} - \mu^{-1}_{k} X_{k} Z  X_{k}
    \end{cases}
    \]

    \STATE Update variables:
    \[
    S_{k+1} = Z, \quad
    X_{k+1} = X_k + D, \quad
    \mu_{k+1} = \eta  \mu_{k}
   \]

    \STATE Check stopping criteria:
    \[
    \max (\|r_p\| , \|r_d\| , \|e_g\| ) \leq \mathrm{tol}_{SDP}
    \]
    \IF{stopping criteria met}
        \RETURN $(X_k, y_k, S_k)$
    \ENDIF

\ENDFOR
\RETURN fail
\end{algorithmic}
\end{algorithm}

Finally, a critical feature of this model is the implementation of CG with variable floating point precision. Precision is one way to improve convergence in dense, random martices that cannot generally benefit from preconditioning. The MPFR Library allows to perform floating point operations at arbitrary machine precision, and it can be implemented as a software layer on top of native $Float64$ architectures, albeit at the expense of a substantial slowdown in execution time \cite{fousse07}. A processor capable of implementing native variable precision arithmetic, the execution time would be much lower. In the appendices we provide estimates of execution times for such a processor. 

This evokes a recurrent theme in scientific computing, namely the distinction between memory-limited and compute-limited applications. Many applications, such as Monte Carlo sampling, sparse matrix-vector multiplication, solving differential equations by finite methods among others, require extensive memory access beyond the cache level, which implies that the execution time and energy dissipation is dominated by shuttling data from memory to CPUs. Scientific computing, therefore, lives typically towards the left of the roofline, in the memory intensive region (see Appendix). So do convex relaxations of combinatorial optimisation problems, in which matrix inversion plays a vital role. 

We expect the proposed approach to be useful for large graphs or complex hard constraints. CG rapidly becomes the only option when matrices are dense and large, and exact matrix inversion is not necessary \cite{zanetti23}.

\section{Results}

Random graphs serve as important test cases for optimisation algorithms. We used the graphs from \emph{Stanford's ``Gset Dataset"}, for which the best-known cuts are publicly available. The graphs used were $[(G17, G19),(G26, G27),(G55, G56),(G63, G64)]$, of with $800, 2000, 5000$ and $7000$ vertices respectively, and a variable number of edges. The adjacency marix of the first (second) graphs in each tuple has positive (and negative) weights.

As explained in previous sections, as the SDP trajectory progresses, computing the Newton step becomes harder as a result of rank-deficiency of tentative solutions. As explained before, preconditioning techniques are not expected to help in problems with dense, unstructured matrices, as is the case of the KKT matrix in the Newton step.s However, we do not rule out finding a preconditioner that will help improve convergence, and leave it for futher wrok. This allows us to assess the effect of extended precision arithmetic exclusively.

\subsection{Methodology}

\begin{figure}[t!]
  \includegraphics[width=\linewidth]{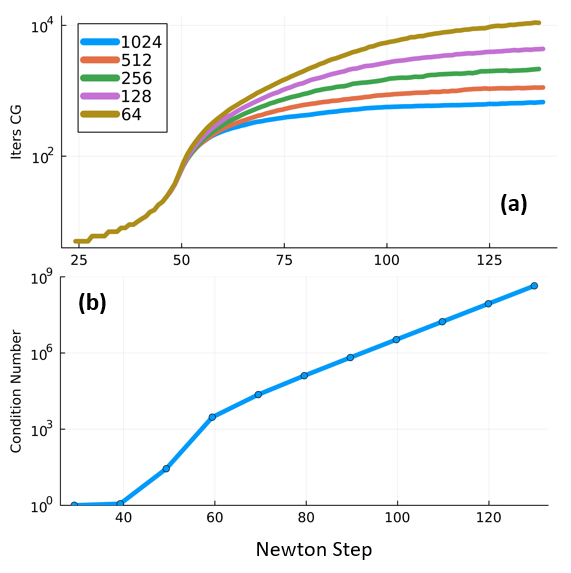}
  \caption{ {\bf (a)} Iterations of CG subroutine vs. iterations of the IPM for $N=5000$. Increasing precision reduces the amount of ``redundant searches" in the Krylov subspace and the matrix is inverted after a smaller number of iterations.{\bf (b)} Condition number $\kappa$ vs. iterations of the IPM. As explained in the appendices, the matrix becomes rank-deficient as the IPM progresses. As a result, the conditioning of the system of equations for matrix inversion increases with the number of iterations. (All data is from graph G55 in the Gset database).} 
\label{fig:itersSDP} 
\end{figure}

\begin{figure}[t!]
  \includegraphics[width=\linewidth]{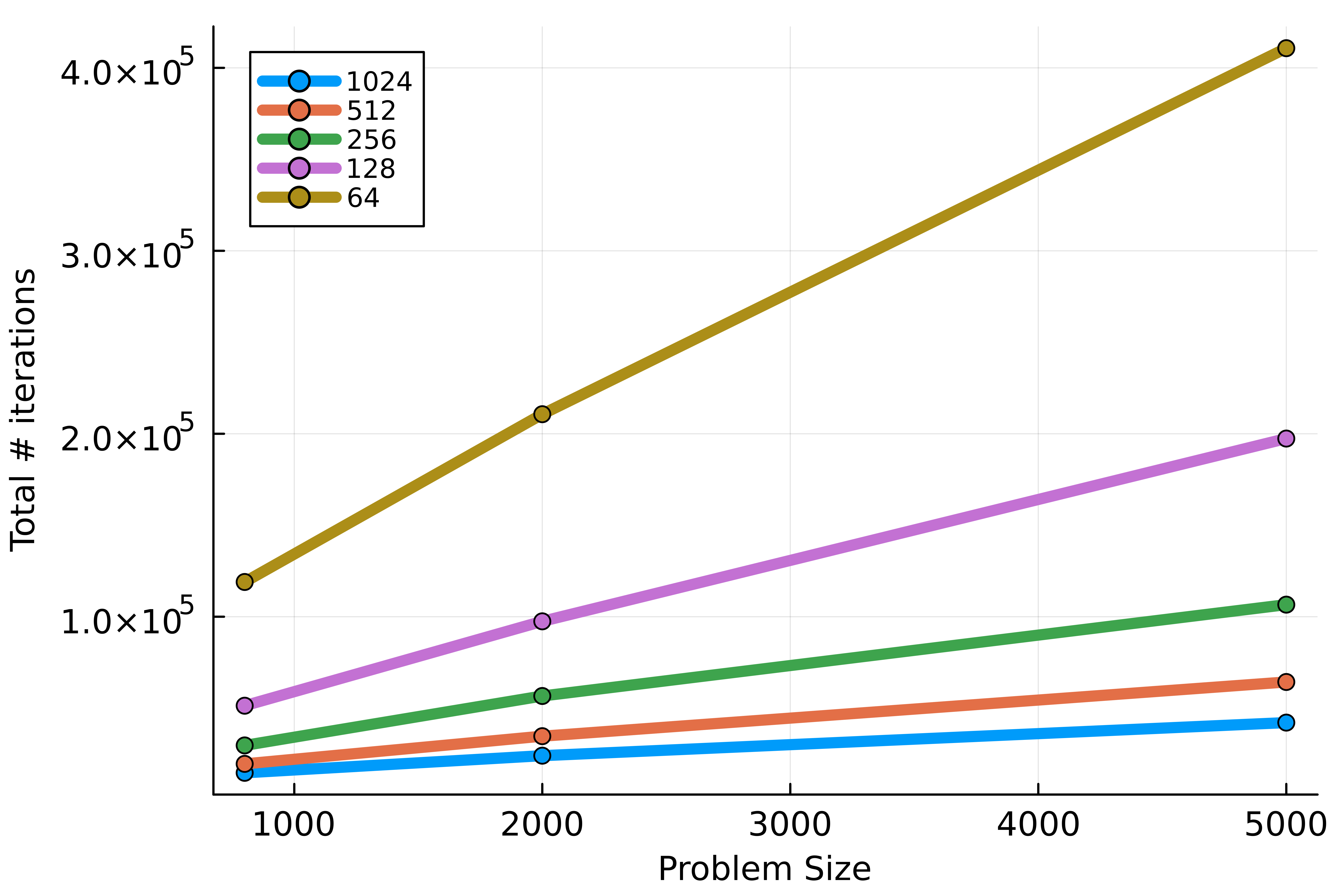}
  \caption{Relative improvement vs. problem size. We integrated the amount of iterations for all floating point precisions considered in this work (1024 bits to 64 bits). This shows that the expected speedup increases with the size of the problem, the slope in the increase of total number of iterations is larger in low precisions compared to that in high precision.  
} 
\label{fig:improvement} 
\end{figure}

We measured the number of needed CG iterations in order to compute the Newton steps along the central path at different precisions, and found that higher floating point precision results in a significant reduction of the number of needed iterations (a theoretical explanation is provided in the appendix). We can link this directly with an explosion of the condition number (see Fig.~\ref{fig:itersSDP}{\bf (b)}). The reduction in the number of iterations becomes more significant with problem size, as shown in Fig.\ref{fig:improvement}).

An important caveat is that each iteration at extended precision takes longer due to the fact that extended precision is being simulated with MPFR \cite{fousse07} on commodity hardware. A processor designed to handle variable precision natively \cite{Guthmuller24} could therefore provide a substantial improvement in these kinds of problems, as discussed below.

We also compared the average value for the best cuts obtain with the GW algorithm to the best cuts available publicly (see Fig. \ref{fig:improvement}). Although it is not surprising that the GW underperforms when compared with the simulated bifurcation machine, we assess that the performance of GW remains constant as the size of the problem increases, as expected from the geometric arguments made to derive the approximation ratio.

\begin{figure}[b!]
  \includegraphics[width=\linewidth]{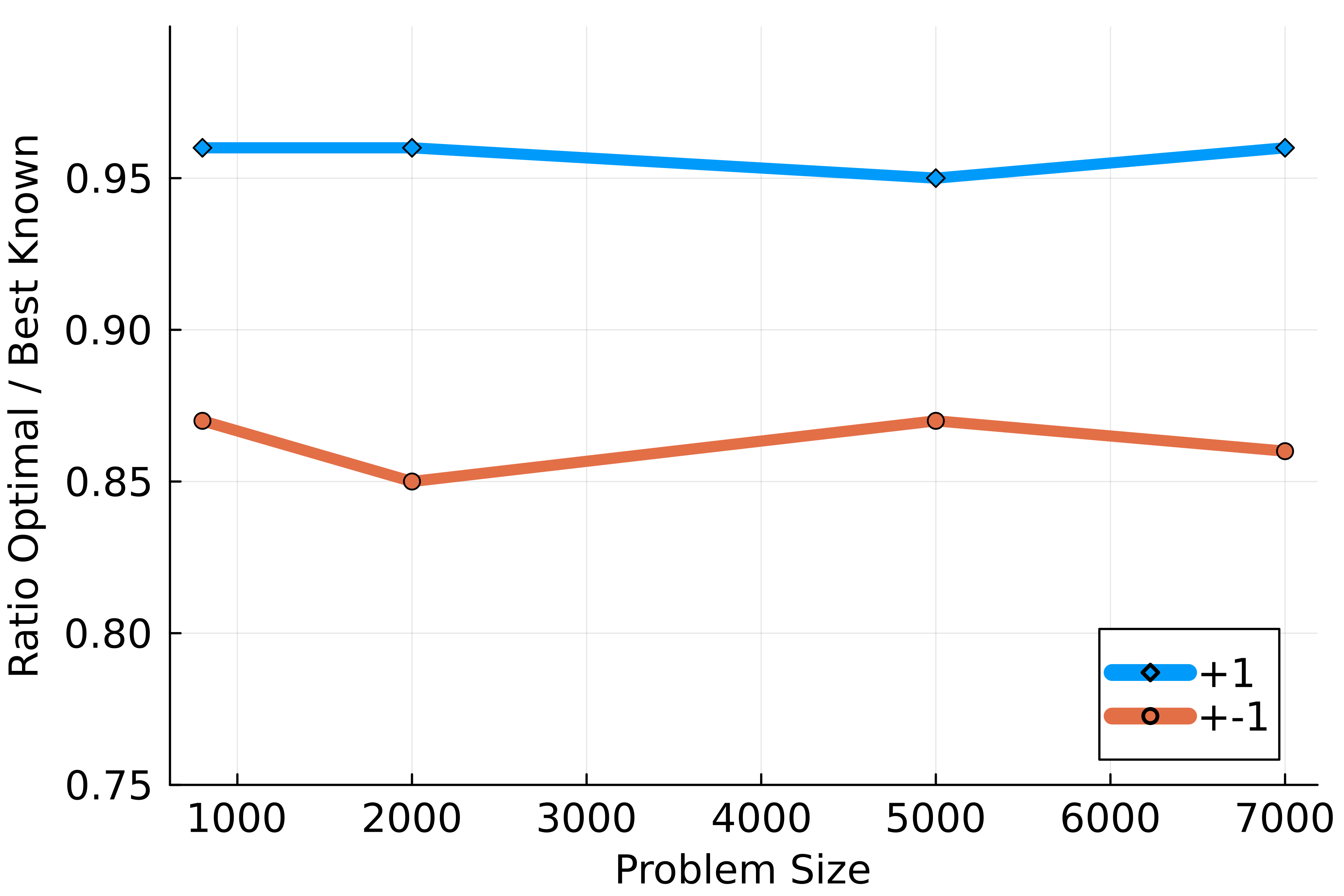}
  \caption{ Ratio between the attained optimal value and the best known cut vs. problem size. The values are averaged over 10 realisations of the random hyperplane separation. Despite providing worst-case guarantees, the GW algorithm becomes gradually competitive with local search heuristics which only provide average-case guarantees. For signed edges, performance is degraded. 
} 
\label{fig:GWfactor} 
\end{figure}

\subsection{Hardware Acceleration Estimations}

We developed a RISC-V based hardware accelerator \cite{Guthmuller24} for extended precision computing.
This RISC-V processor has been fitted with a variable and extended precision floating point unit and a corresponding instruction set extension.
It supports up to 512-bit floating point precision with performance depending on input and output precision.
We have validated and measured this hardware implementation on real chips \cite{Fuguet24,Guthmuller25}.
Our newest version of this processor, called VXP, has been optimized to reduce the overhead associated with increased precision: we improved instructions latency and throughput at high precision and introduced dedicated hardware for sparse matrix support to reduce accesses to external memory.

Even though our accelerator's ASIC implementations can run the CG routine that lies at heart of the IPMs in the large problem limit, they embed 8 cores at most, and thus lack enough parallelism to really speedup computation over the MPFR emulation used previously.
However, we can easily extrapolate performance on a hypothetical multicore high performance implementation that would exploit the intrinsic parallelism of linear algebra routines used in the kernel.
To do so, we developed a high-level simulator of private and shared caches behaviour, to estimate the number of accesses to external memory.

Our simulation shows that for the CG routine applied on a dense 64-bit input matrix, the number of external memory accesses is almost independent of the precision of internal vectors \cite{Guthmuller24}.
It can easily be explained by noting that most of these memory accesses are issued during the matrix-vector multiplication and are caused by the streaming of the matrix elements.

To estimate the number of cores we need to saturate memory, we use the roofline model \cite{Williaws09}(see Appendix).
At 512-bit precision, our accelerator sends one 64-bit memory load every four cycles during dense matrix-vector multiplication.
Assuming our hardware accelerator is connected to two HBM3E stacks with a resulting memory bandwidth of 2.4 TB/s \cite{Kim24}, we would thus need around 600 cores to be limited by memory accesses.
Given previous hardware implementations of our processor in 7 nm technology \cite{Guthmuller24}, the resulting ASIC would have a silicon area of around 400 mm² .

\begin{figure}[b!]
  \includegraphics[width=\linewidth]{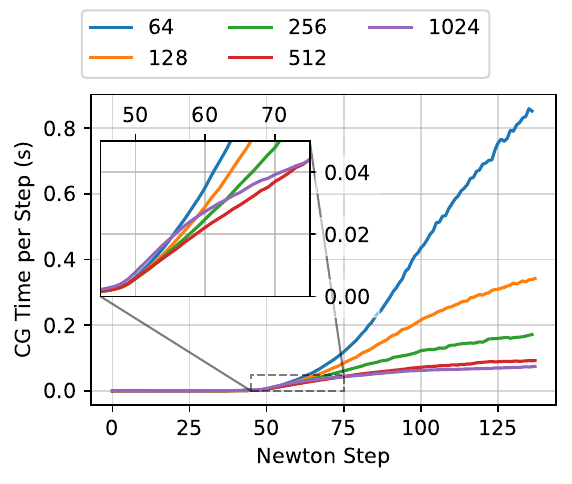}
  \caption{Time spent in CG algorithm at each Newton step for G55 problem (5000x5000).} 
\label{fig:G55_time} 
\end{figure}

Finally, we get the time needed by such an accelerator to execute a CG iteration by dividing the number of memory accesses by the 2.4 TB/s memory bandwidth.
Our hardware does not currently support 1024-bit precision, but adding support for it is feasible and we estimate that it would result in a 20\% execution time penalty compared to 512-bit precision. Fig.~\ref{fig:G55_time} shows the execution time of each CG execution at each Newton step for different precisions.
The figure includes a focus on a small region where CG execution time at higher precision is progressively lower than execution time at lower precision.
This observation leads us to follow an adaptive scheme where the best precision is used at each time step, leading to better performance and lower power consumption.
This could be achieved using heuristics such as running the CG at higher precision every few Newton steps and switch if the gain is significant.

\begin{figure}[b!]
  \includegraphics[width=\linewidth]{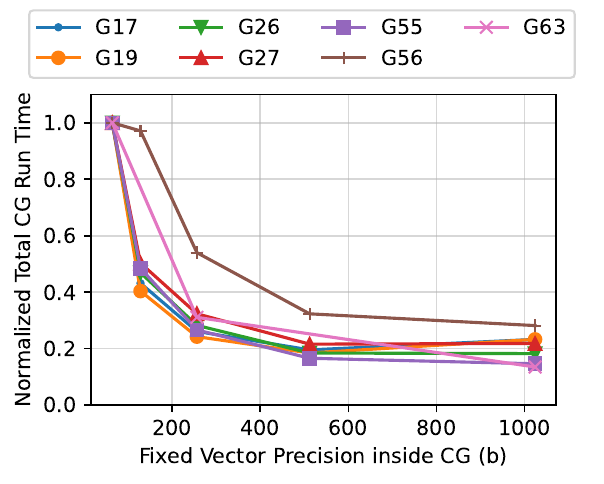}
  \caption{Total time spent in CG for all problems and precisions, normalised to 64b precision.} 
\label{fig:total_time} 
\end{figure}

Fig.~\ref{fig:total_time} shows the relative total time spent in CG for the full GW algorithm, normalized to the 64-bit precision to demonstrate the benefit of extended precision.
Generally, 1024-bit precision is the fastest and problems having only positive weights benefit the most from extended precision.
Table~\ref{tab:cg_total_time} summarises the total time spent in CG using the best precision at each step.
As expected, execution time grows very rapidly in the  problem size.
However, thanks to the iterative nature of CG, the memory footprint is mostly limited to storing the input matrix.

Hardware support for extended precision reduces convergence time by up to 10$\times$, compared to that of 64-bit precision.
Crucially, this reduction also seems to be increasing with problem size, and suggests that gains are set to increase when considering very large graphs.
Our proposed variable precision scheme also reduces execution time by up to 27\% compared to fixed 1024-bit precision, and should also reduce significantly the power consumption at the start of the GW algorithm.

\begin{table}[tbp]
    \caption{Total Adaptive CG execution time using the best precision at each Newton step, along with relative gains compared to 64-bit and 1024-bit precisions}
    \label{tab:cg_total_time}
    \begin{center}
        \small
        \begin{tabular}{l|l|l|l}
        \textbf{Problem} & \textbf{CG time (s)} & \textbf{Gain vs 64b} & \textbf{Gain vs 1024b} \\
        \hline
        G17 & 0.05 & 5.1$\times$ & +20\% \\
        G19 & 0.04 & 5.5$\times$ & +27\% \\
        G26 & 0.46 & 5.7$\times$ & +5\% \\
        G27 & 0.42 & 4.9$\times$ & +7\% \\
        G55 & 4.58 & 7$\times$   & +2\% \\
        G56 & 14.4 & 3.6$\times$ & +2\% \\
        G63 & 75.1 & 10$\times$ & +1\% \\
    \end{tabular}
    \end{center}
\end{table}

\section{Conclusion}

We explored how using extended precision helps to improve the time to solution in a convex semidefinite relaxation of the combinatorial problem Max-Cut, which is frequently used as a benchmark for quantum and classical optimisers. 

As the optimisation advances, calculating the next step becomes harder and harder as a result of rank-deficiency of the tentative solution to the convex relaxation. When using indirect methods like Conjugate Gradient, which have lower complexity than direct methods and are therefore used in very large problems, we see that increasing the internal working precision reduces the time to solution by a factor that appears to increase with the system size.

Whereas the fact that extending the precision can improve convergence times in Krylov methods was already a well-known result, this work aims at showing how this can be incorporated as a subroutine in convex optimisation, namely in Interior Point Methods that are used to solve SDP programs. 

When comparing direct approaches to matrix inversion implemented in Julia, which makes calls to an optimsed BLAS library and found that in the case of CG in $Float64$ precision, which runs without MPFR, the speed of execution was comparable.

Another goal of this work was to estimate the acceleration that would be achievable in a hypothetical hardware accelerator that allows for native use of extended precision. We found that, given a realistic hardware budget, we could build a hardware accelerator solving the Newton steps up to 10 times faster with a constrained memory bandwidth in our experiments, thanks to high precision computing. 

\section{Acknowledgements}

We would like to thank Y. Durand for enriching and fun conversations on numerical algebra and for guiding us through the fascinating world of numerical stability.

\bibliographystyle{unsrt}

\pagebreak

\section*{Appendix}

\subsection*{A. Primal-Dual Barrier Method for Semidefinite Programming}

By introducing the barrier $\mu$,  the original SDP cost function is modified with a log-determinant barrier function \cite{forsgren02}:

\[
\begin{aligned}
\min_{X} &\quad \mathrm{Tr}(C X)  - \mu \log\det(X)\\
\text{subject to}&\quad \Tr(A_i X) = 1 \\
&\quad X \succ 0 
\end{aligned}
\]

In our code, the iterates follow a so-called ``$\beta$-approximate central path" given by the relaxed KKT conditions:

\[
\begin{aligned}
&\quad  \Tr(A_i X)  = 1 \\
&\quad X = L^T L \\
&\quad\sum y_i A_i + S = C \\
&\quad \| I - \mu^{-1} L^T S L \| \leq \beta\\
&\quad X \succ 0, \quad S \succ 0
\end{aligned}
\]

As the barrier parameter shrinks and vanishes, $\mu \to 0 $, the iterates $X(\mu), S(\mu) $ approach the optimal solution, which typically lies at the boundary of the positive semidefinite cone, i.e., $X \succeq 0 $ but not necessarily $X \succ 0 $. In other words, as the tentative solution approaches this boundary, the matrix $X $ approaches a singular matrix:

\[
\mu \to 0 \quad \Rightarrow \quad \det(X) \to 0  
\]

Assuming that the tentative solution is in the central path, it is possible to obtain the Newton step $\Delta$ from a quadratic approximation of the cost function:

\[
\begin{aligned}
\min_{\Delta} &\quad \mathrm{Tr}((C  - \mu X^{-1})\Delta) \\
&\quad + \frac{\mu}{2}\Tr(X^{-1}\Delta X^{-1}\Delta )\\
\text{subject to}&\quad \Tr(A_i \Delta) = 0 
\end{aligned}
\]

The first term in the cost function corresponds to the gradient of log-determinant barrier cost function, and the second one is proportional to the Hessian. It is noteworthy that the Hessian (the curvature of the optimisation problem for the Newton step) is inversely proportional to an inverse power of the tentative solution, which explains why CG struggles as $X$ becomes rank-deficient.

Since the condition number of $X$ is:

\[
\kappa(X) = \frac{\lambda_{\max}(X)}{\lambda_{\min}(X)}
\]

As $\lambda_{\min}(X) \to 0 $, the condition number $\kappa(X) \to \infty $, which means that the KKT equations are ill-conditioned along directiongs in which the tentative solution has vanishing eigenvalues. Conversely, small numerical errors in the Newton system lead to large errors in $\Delta $, which, if too big, can result sometimes in leaving the cone of semipositive definite matrices.

\subsection*{B. The Conjugate Gradient Algorithm}

The CG method is based on minimising the quadratic functional:

\[
f(x) = \frac{1}{2} x^T A x - b^T x
\]

where $A \in \mathbb{R}^{n \times n} $ is a symmetric, positive matrix. The minumum of this function is located at $x^*$ such that $\nabla f(x) = 0$, giving $Ax^* = b$. Instead of minimising $f(x) $ over all of $\mathbb{R}^n $, CG restricts the minimisation to take place within the \emph{Krylov subspace}:

\[
\mathcal{K}_k(A, b) = \text{span} \{b, Ab, A^2 b, \dots A^{k-1} b \}
\]

CG is suited for large sparse systems since it does not require storing the entire matrix or its inverse. At each iteration, the tentative solution is updated as:

\[
x_{k+1} = x_k + \alpha_k p_k
\]
where $p_k $ is the search direction and $\alpha_k $ is a scalar chosen to minimize the error along $p_k $. This is implicitly building a tridiagonal Lanczos matrix.

In exact arithmetic, the CG method generates search directions $\{p_k\} $ that are A-conjugate:
\[
p_i^T A p_j = 0 \quad \text{for } i \ne j,
\]
and residuals $\{r_k\} $ that are mutually orthogonal:
\[
r_i^T r_j = 0 \quad \text{for } i \ne j.
\]

The CG method converges in at most $n $ iterations in exact arithmetic, typically much faster depending on the condition number $\kappa(A)$. The error decreases accordingly to the expression:

\[
\delta^{(\textrm{exact arithmetic})}_k   \leq \left(\frac{\sqrt{\kappa}-1}{\sqrt{\kappa}+1}\right)^k
\]

At finite precision, the A-conjugacy property is lost progessively, so the search proceeds along redundant directions. The cummulated round-off error at iteration $k$ is:

\[
\delta^{(\textrm{finite precision})}_k \approx \epsilon_{\textrm{precision}} k \sqrt \kappa
\]

Rounding errors accumulate over iterations, which leads to a loss of orthogonality among residuals and a loss of conjugacy among directions. These effects are especially problematic when the matrix $A $ is ill-conditioned In this case, small numerical errors can be amplified, causing search directions to become very nearly parallel (a bit like steepest descent), which in turn slows convergence and the method may stall \cite{greenbaum97}.

The relative error, measured in terms of residues, can be understood as being bounded by two different terms, one decreasing with the iteration and related to exact arithmetic, and the other one increasing as a consequence of finite precision:

\[
\delta^{(\textrm{TOTAL})}_k = \delta^{(\textrm{exact arithmetic})}_k + \delta^{(\textrm{finite precision})}_k
\]

The first one decreases with the iteration, as expected. The other one increases with the condition number and gets worse as CG increases. This implies the existence of a sweet-spot beyond which adding more iterations to CG becomes ineffective.

\subsection*{C. MPFR and Hardware Considerations}
In this work, extended precision was implemented through two means: a software library, MPFR\cite{fousse07}, and a hardware implementation based on a dedicated accelerator, the VRP.

MPFR is a software library that enables the calculation on floating-point numbers with arbitrary precision, down to the bit level, beyond the current hardware limitations (64 bits). Floating-point numbers in MPFR are represented using a dedicated data structure, \texttt{mpfr\_t}, which carries the information related to the number and is used by specific functions to perform high-precision arithmetic operations.

The \texttt{mpfr\_t} structure carries the following information: precision: the number of bits representing the mantissa of the floating-point number, exponent: the value of the exponent, as a power of 2, of the represented number, mantissa: the value of the mantissa of the represented number, sign: the sign of the number, either positive or negative.

A number in MPFR format can then be passed as a parameter to the arithmetic functions available in the library, such as addition (\texttt{mpfr\_add}), subtraction (\texttt{mpfr\_sub}), multiplication (\texttt{mpfr\_mul}), division (\texttt{mpfr\_div}), etc. 
In addition to these functions, MPFR also provides functions for managing the memory associated with these structures, display functions, and type conversion functions (to and from standard floating-point numbers such as double and float).

\begin{lstlisting}
#include <mpfr.h>

int main() {
    mpfr_t x;
    /* Initialize x with a 
     * precision of 53 bits */
    mpfr_init2(x, 53);

    /* Usage of x
     * Assign the value 123 to x */
    mpfr_set_ui(x, 123, MPFR_RNDN);
    
    /* Add x to itself and store 
     * the result in x           */
    mpfr_add(x, x, x, MPFR_RNDN);

    /* Free the memory */
    mpfr_clear(x);

    return 0;
}
\end{lstlisting}

However, software emulation of extended precision computing being quite inefficient motivated our work on an hardware implementation of an extended precision accelerator.
As we wanted to ease the programming of this accelerator, we chose to implement extended computing instructions in a 64b RISC-V general purpose processor.
While implementing efficient arithmetic operators is of prime importance, typical scientific computing kernels are generally limited by memory bandwidth.
As external memory access is slow (from hundreds to thousands of cycles) processors implement multiple level of hardware caches.
The first level is the closest to the core and accessible in few cycles at very high throughput, while farther ones latency reaches tens of cycles.

Processor caches improve performance by exploiting two behaviours exhibited by applications:
\begin{enumerate}[topsep=0pt,itemsep=-1ex,partopsep=1ex,parsep=1ex]
    \item Spatial locality: successive memory accesses are generally done to close locations, most often consecutive memory addresses.
    \item Temporal locality: a memory location has a high chance to be accessed multiple times during the execution of the program.
\end{enumerate}
To exploit spatial locality, caches store blocks of data, called \textit{cache lines} or \textit{cache blocks}, instead of a single variable.
The VXP accelerator first level caches implements 64B cache lines, which is typical for general purpose processors.
To exploit temporal locality, caches try to keep most accessed cache lines longer in the cache.
This process is called \textit{cache replacement policy} and is critical to reduce accesses to higher level caches or external memory.

Finally, modern processor try to fill the cache not only at the moment they are accessed, but also by anticipation.
This behaviour, called \textit{prefetching}, is essential to mask memory latency and can either be explicitly programmed by the application (supposing the application knows its future memory accesses) or by the hardware itself doing guesses.
The VXP only provides explicit prefetching, with additional hardware support for sparse matrices which generate difficult-to-predict memory accesses.

\begin{figure}[h!]
  \includegraphics[width=0.8\linewidth]{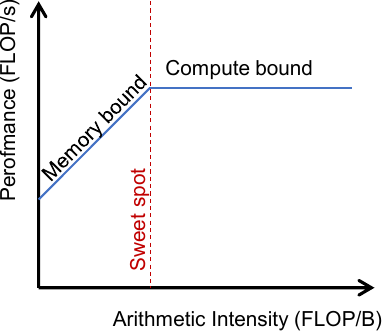}
  \caption{ The Roofline model provides a visual representation of the computational peak performance limits imposed by memory bandwidth and processing capacity. This model characterizes algorithms based on their numerical intensity, i.e., the ratio of floating-point operations to memory calls. 
} 
\label{fig:roofline} 
\end{figure}

When designing an hardware accelerator, a system architect has to dimension the number of cores, the cache hierarchy and the external memory bandwidth given the properties of the target application.
One of the most important property is the arithmetic intensity of the application: how many computation, expressed in Floating Point OPerations (FLOP), for each byte accessed in memory.
The arithmetic intensity can than be used to obtain to first order the performance of the application on a given hardware thanks to the roofline model \cite{Williaws09} shown in Fig.\ref{fig:roofline}.
This model uses the peak FLOP/s and memory bandwidth of the underlying hardware to estimate the maximum performance that could by obtained given the arithmetic intensity of the application.
Even if this is a very simplified model it is helpful to either adapt the hardware to the application or vice versa.

\end{document}